\documentclass{article}
\usepackage{ijcai19}

\usepackage{times}
\usepackage{soul}
\usepackage{url}
\usepackage[hidelinks]{hyperref}
\usepackage[utf8]{inputenc}
\usepackage[justification=justified, skip=2pt]{caption}
\usepackage{graphicx}
\usepackage{amsmath}
\usepackage{booktabs}
\usepackage{adjustbox}
\usepackage{subcaption}
\usepackage{bbm}
\usepackage{color}
\usepackage{multirow}
\usepackage{enumitem}
\usepackage{algorithm} 
\usepackage{algorithmic}
\usepackage{enumerate}
\usepackage{multirow}
\usepackage{stmaryrd}
\usepackage{mathrsfs} % needed for mathscr
\usepackage{color}
\usepackage{amsfonts}
\usepackage{balance}

\usepackage[para,online,flushleft]{threeparttable}
\newcommand{\etal}{\textit{et al}. }
\newcommand{\etc}{\textit{etc}.}
\newcommand{\eg}{\textit{e.g.}}
\newcommand{\ie}{\emph{i.e.}}

\usepackage{soul}   % for strikethrough
\setlist[itemize]{leftmargin=*}
\usepackage{xcolor}
\usepackage{color}

\newcommand{\citet}[1]{\citeauthor{#1}~\shortcite{#1}}

\renewcommand{\arraystretch}{0.85}

\makeatletter
\g@addto@macro\normalsize{%
  \abovedisplayskip 3pt plus 2pt minus 3pt%
  \belowdisplayskip \abovedisplayskip
  \abovedisplayshortskip 3pt plus 2pt minus 3pt%
  \belowdisplayshortskip 3pt plus 2pt minus 3pt%
}

\makeatother

\title{Neural Re-ranking in Multi-stage Recommender Systems: A Review}

\author{
Weiwen Liu$^1$\and
Yunjia Xi$^2$\and
Jiarui Qin$^2$\and
Fei Sun$^3$\and
Bo Chen$^1$\and
Weinan Zhang$^2$\and \\
Rui Zhang$^4$\and
Ruiming Tang$^1$
\\
\affiliations
$^1$Huawei Noah's Ark Lab $\quad$
$^2$Shanghai Jiao Tong University \\
$^3$Institute of Computing Technology, Chinese Academy of Sciences$\quad$
$^4$ ruizhang.info\\
\emails
\{liuweiwen8,chenbo116,tangruiming\}@huawei.com, \{xiyunjia,qinjr96,wnzhang\}@sjtu.edu.com,\,rayteam@yeah.net
}

\begin{document}

\maketitle

\begin{abstract}
As the final stage of the multi-stage recommender system (MRS), re-ranking directly affects users’ experience and satisfaction by rearranging the input ranking lists, and thereby plays a critical role in MRS. With the advances in deep learning, neural re-ranking has become a trending topic and been widely applied in industrial applications. This review aims at integrating re-ranking algorithms into a broader picture, and paving ways for more comprehensive solutions for future research. For this purpose, we first present a taxonomy of current methods on neural re-ranking. Then we give a description of these methods along with the historic development according to their objectives. The network structure, personalization, and complexity are also discussed and compared. Next, we provide benchmarks of the major neural re-ranking models and quantitatively analyze their re-ranking performance. Finally, the review concludes with a discussion on future prospects of this field. A list of papers discussed in this review, the benchmark datasets, our re-ranking library \texttt{LibRerank}, and detailed parameter settings are publicly available\footnote{https://github.com/LibRerank-Community/LibRerank}.
\end{abstract}

\section{Introduction}
Multi-stage Recommender Systems (MRS) are widely adopted by many of today's largest online platforms, including Google \cite{seq2slate}, YouTube \cite{wilhelm2018practical}, LinkedIn \cite{geyik2019fairness}, and Taobao \cite{prm}. MRS is a natural solution to the computational limits in practical recommendation applications, where the numbers of users and items grow into billions. 
The recommendation task is split into multiple steps in MRS---each step narrows down the relevant items with a slower but more accurate model \cite{hron2021component}, to guarantee low response latency. A common structure for MRS consists of three stages in general: candidates generation (\emph{a.k.a.,} recall or matching), ranking, and re-ranking. The system firstly generates candidates from a large pool of items. Then these candidates are scored and ranked in the ranking stage. 
Finally, the system conducts re-ranking on the top candidates based on certain rules or objectives to further improve the recommendation results. Specifically, the re-ranking stage takes as input the initial ranking list from the ranking stage, and outputs a re-ordered list by considering the listwise context (cross-item interactions). 
Whether a user is interested in an item is not only determined by the item itself, but also by other items placed in the same list (\ie, the listwise context) \cite{prm}. 
Thus, a key technical challenge is to model the listwise context in re-ranking.

\begin{figure}[t]
    \centering
    \includegraphics[width=\columnwidth]{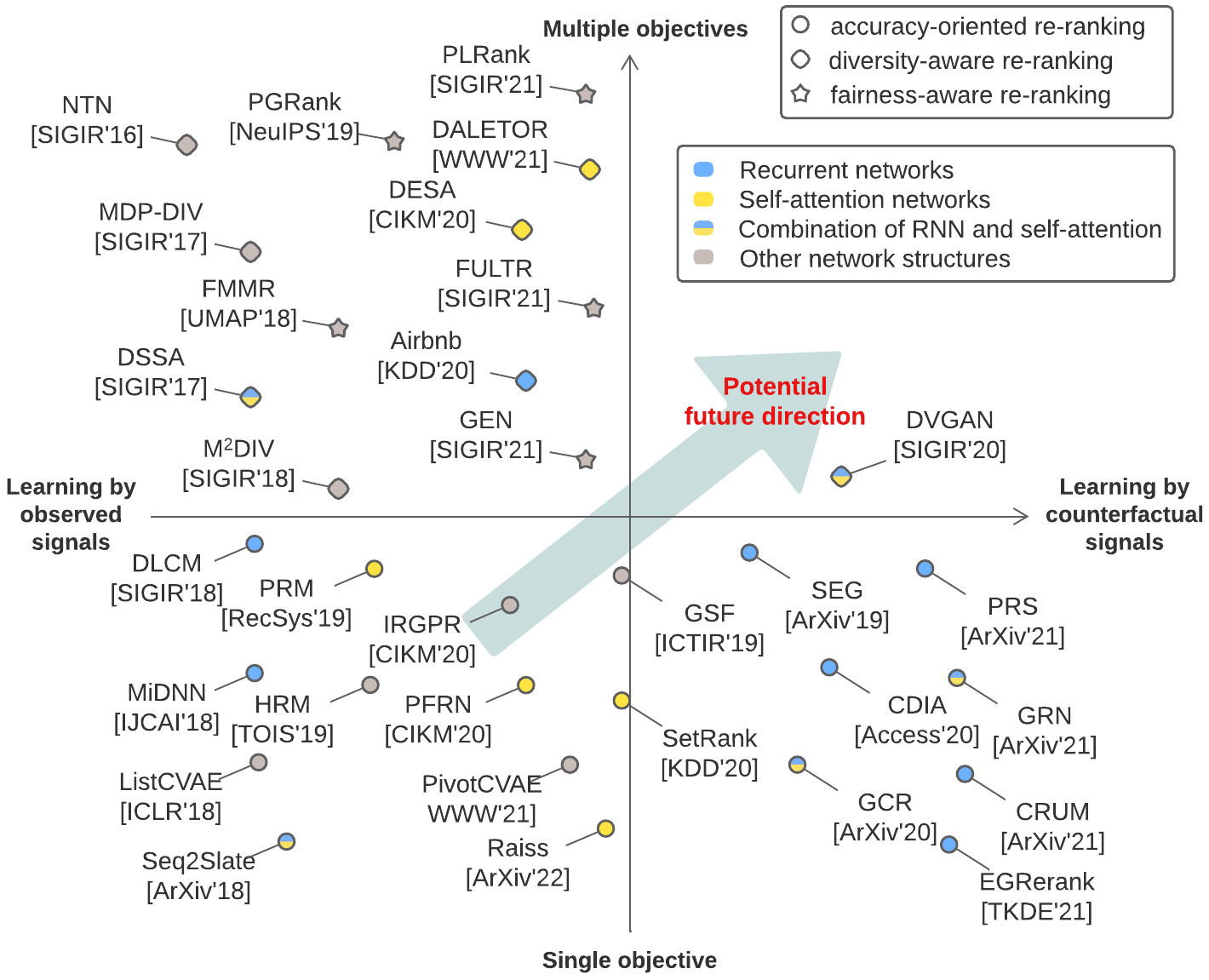}
    \caption{Four quadrants of neural re-ranking models. Shapes denote different objectives, and colors represent the major network structures for re-ranking.}
    \label{fig:orgchart}
\end{figure}

Re-ranking dates back to Carbonell's work \shortcite{carbonell1998use}, which greedily adds items to the list with maximal marginal relevance. With deep neural networks led to exciting breakthroughs in various fields \cite{goodfellow2016deep,goldberg2017neural}, re-ranking methods evolved to the recent deep neural architectures. 
Neural re-ranking models get rid of the hand-crafted features and benefit from the automatic learning of the listwise context, due to the universal approximation property of neural networks \cite{cybenko1989approximation}. Therefore, applying neural networks for re-ranking is the main focus in both academia and industry in recent years. This paper provides a first review on neural re-ranking for recommendation.

% \begin{figure}[t]
%     \centering
%     \includegraphics[width=\columnwidth]{imgs/mrs.pdf}
%     \caption{A multi-stage recommender system.}
%     \label{fig:mrs}
% \end{figure}

\subsection{A Taxonomy} 
We differentiate neural re-ranking models by objectives (single accuracy objective or multiple objectives) and the supervision signals (observed signals or counterfactual signals). The resulting four quadrants are outlined in Fig.~\ref{fig:orgchart}.

Considering the objectives, most studies focus on the single accuracy objective, as accurately predicting users' interests is the foundation of recommender systems. While beyond accuracy, several other objectives are also desired from a re-ranking model like diversity or fairness, thus leading to recent work on how to optimize multiple objectives in re-ranking and better manage the tradeoff between them.

Another important factor that separates different re-ranking models is the supervision signals for relevance. Most work is directly trained by the displayed initial ranking lists and the corresponding observed labels. Some other work, however, points out that the relevance of each item depends on the listwise context, and different permutations of the input list yield different relevance labels \cite{prs}. The supervision signals are therefore provided by an extra evaluator on the unobserved counterfactual permutations that have not been actually displayed to the user, to model the listwise context under different permutations.

From Fig.~\ref{fig:orgchart}, we observe the following development characteristics of existing neural re-ranking work: (i) Most studies seek to purely enhance accuracy with a single accuracy objective, while diversity/fairness-aware methods with multi-objectives have been relatively less explored. (ii) Self-attention \cite{vaswani2017attention} or a combination of RNN \cite{hochreiter1997long} and the attention have become popular network structures in re-ranking. (iii) Few works discuss the influence of counterfactual permutations on relevance in multi-objective learning (the first quadrant of Fig.~\ref{fig:orgchart}), which could be a potential research direction. We will elaborate on the details of each method according to our proposed taxonomy in the following sections.

% Note that although some work claims itself to be a ranking model, it requires additional complex listwise context as input, so that is mainly placed at the re-ranking stage in MRS to meet the online latency requisite \cite{hron2021component,chen2017efficient}. We also include and discuss this line of work in this review. 

\section{Neural Re-ranking for Recommendation}
 Neural re-ranking usually aims to construct a multivariate scoring function, whose input is a whole list of items from the initial ranking, to model the listwise context/cross-item interactions \cite{setrank,gsf}. This is in contrast to ranking models where the ranking functions are mostly univariate that take one item at a time, and the correlations between items are only modeled at loss level using pairwise or listwise loss functions \cite{xia2008listwise}. 

% The re-ranking stage takes as input an initial ranking list from the prior ranking stage, and output a refined list.

For a specific user, given the initial list $R$ of $n$ items, and the corresponding supervision signals $Y\in\mathbb R^n$, a neural re-ranking problem is to find the optimal ranking
function $\phi_*$ that maps the input to a list of re-ranking scores as
\begin{equation}\label{eq:rerank_function}
    \phi_*=\arg\min_{\Phi} \sum_{R,Y} \mathcal L(Y,\phi(R))\,,
\end{equation}
where $\mathcal L(\cdot)$ is the loss function.
The major goal of the re-ranking is to optimize accuracy, which is usually measured by ranking metrics like NDCG or MAP. While beyond accuracy, encouraging diversity or fairness of the re-ranking is also one of the critical goals. 

This general formulation of Eq.\eqref{eq:rerank_function} provide another perspective to describe our proposed taxonomy in  Fig.~\ref{fig:orgchart}. The design of the loss function, either is purely accuracy-oriented or a combination of multiple objectives, differentiates re-ranking models into the single objective and the multi-objective ones. On the other hand, whether the supervision signal $Y$ comes from the data log or an evaluator, separates re-ranking models into learning by observed signals or by counterfactual signals. Fig.~\ref{fig:flowchart} shows typical network architectures for re-ranking. Learning by observed signals usually follows a direct architecture, outputting re-ranking scores with listwise context modeling. Whereas learning by counterfactual signals generally adopts a generator-evaluator paradigm---the generator generates re-ranking lists under the guidance of an evaluator, where both the generator and the evaluator attend to the listwise contexts. Later we will introduce different neural re-ranking models according to the four quadrants in detail. 

% Single objective re-ranking is introduced in Section \ref{sect:acc}. Specifically, learning by observed signals and by counterfactual signals are discussed in Section \ref{sect:observed} and Section \ref{sect:counterfactual}, respectively. The multi-objective re-ranking is described in Section \ref{sect:multiple}, followed by some emerging applications in Section \ref{sect:app}.

% \begin{figure}[t]
%     \centering
%     \includegraphics[width=\columnwidth]{imgs/flowchart.pdf}
%     \caption{The flowchart of learning a neural re-ranking model.}
%     \label{fig:flowchart}
% \end{figure}

\section{Single Objective: Accuracy-oriented}\label{sect:acc}

Recommendation accuracy of the re-ranking model is the fundamental goal for MRS, and the evaluation of a re-ranking model is usually the \textit{overall listwise utility} like NDCG or MAP of the re-ranking list.
% Compared to the classification/regression error adopted in the recall or ranking stage, t

According to the supervision signal, we further divide existing re-ranking models into two groups: learning by observed signals and learning by counterfactual signals. Learning by observed signals directly uses the initial ranking list $R$ and the corresponding label $Y$, which is actually displayed to the user and obtained feedback, to train the model. On the contrary, learning by counterfactual signals presumes the item's relevance varies under different permutations---even with the same items, users respond distinctly to different permutations of these items. Therefore, they introduce an additional evaluator to provide signals for counterfactual permutations that have not been actually displayed to the user. Listwise context is thereby estimated on the counterfactual permutations. Below we describe various attempts with their advantages and disadvantages for methods of learning by observed signals and learning by counterfactual signals.

\subsection{Learning by Observed Signals}\label{sect:observed}
\begin{figure}[t]
    \centering
    \includegraphics[width=\columnwidth]{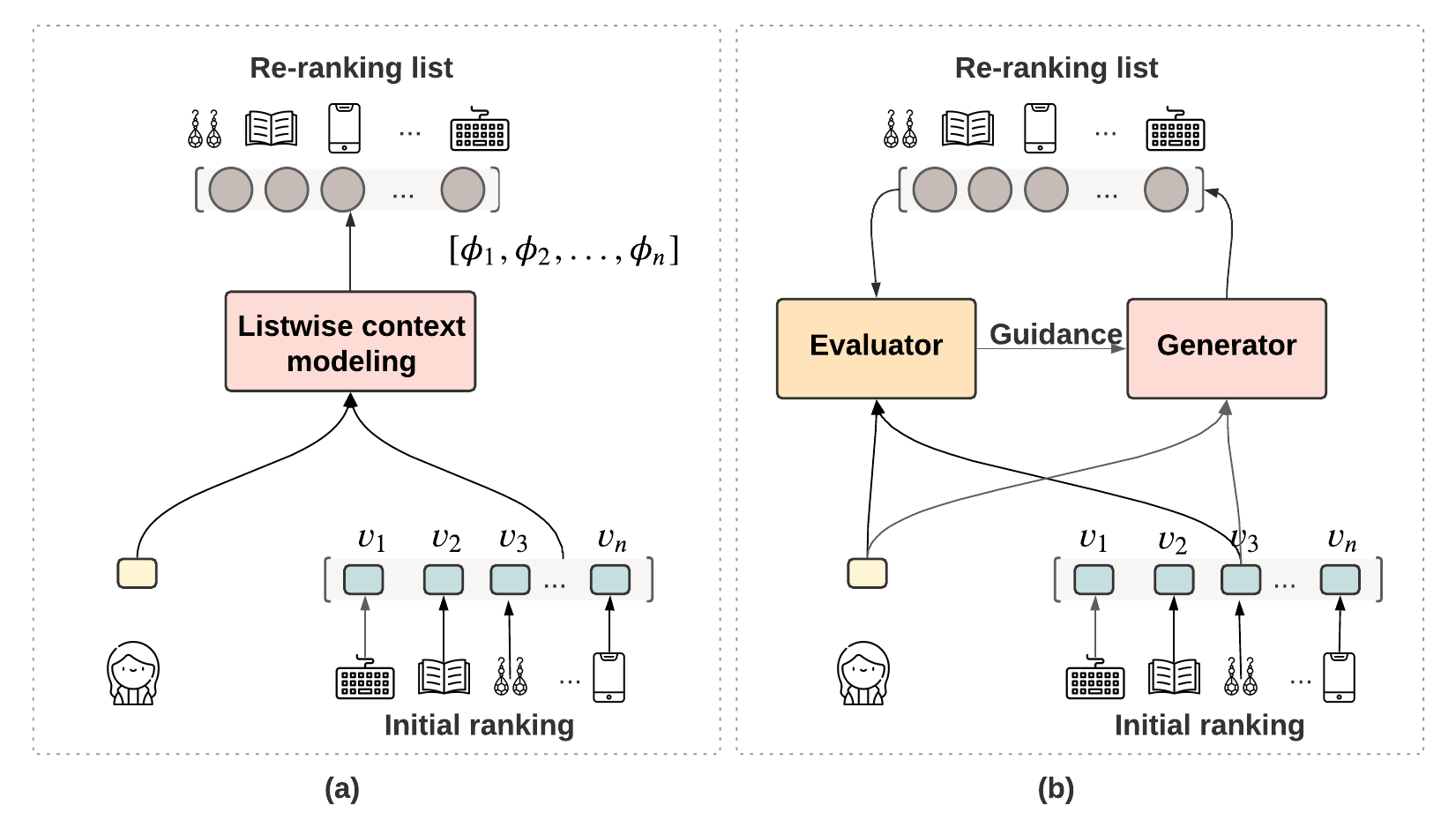}
    \caption{Re-ranking network architectures. (a) A typical neural re-ranking architecture for learning by observed signals. (b) The evaluator-generator paradigm for learning by counterfactual signals.}
    \label{fig:flowchart}
\end{figure}

Learning by observed signals is simple and straightforward. A typical architecture of the re-ranking model for learning by observed signals can be outlined as in Fig.~\ref{fig:flowchart}(a), which firstly embeds user and item features into low-dimensional dense vectors, and then extracts cross-item interactions by the listwise context modeling to generate the re-ranking scores. The observed labels are actual feedback from users, and thus are less noisy and easier to train. Moreover, the initial list provides strong signals for items' relevance estimated by previous ranking models. Several existing studies have shown the effectiveness of directly learning by observed signals \cite{dlcm,prm,setrank,feng2021revisit}. By the network structure adopted to model the listwise context, we further classify the existing methods into \textit{recurrent listwise modeling} with recurrent neural networks (RNN), \textit{attentive listwise modeling} with self-attention, and \textit{others} like multi-layer perceptrons (MLP), graph neural networks (GNN), \etc, where different network structures are also plotted in Fig.~\ref{fig:orgchart} by different colors.

\smallskip
\noindent
\textbf{Recurrent listwise modeling.} As one of the earliest neural re-ranking methods for improving accuracy, DLCM \cite{dlcm} uses gated recurrent units (GRU) to sequentially encode the top-ranked items with their feature vectors. The recurrent unit combines the information for the current item with previous items, which naturally captures the sequential dependencies among items and the positional effect of the initial list.
MiDNN \cite{midnn} also applies recurrent networks, the long-short term memory (LSTM), with a global feature extension method to capture cross-item influences. It formulates the re-ranking as a sequence generation problem, and sequentially selects the next items with beam search to conform to the users' browsing habit. 
Seq2Slate \cite{seq2slate} extends MiDNN by adopting a more flexible pointer network to solve the re-ranking problem. The pointer network produces the next item with an attention mechanism, attending to the items in the initial list. 

\smallskip
\noindent
\textbf{Attentive listwise modeling.} 
Lately, inspired by the success of the self-attention architecture used in natural language processing \cite{vaswani2017attention}, several re-ranking models that apply the multi-head self-attention are proposed. Compared to RNN, the self-attention mechanism directly models the interactions between any pair of candidate items without degradation over the encoding distance. 
PRM \cite{prm} is a generally straightforward adaptation of the self-attention structure, which is a stack of multiple blocks of self-attention layers and feed-forward networks with position embeddings of the initial list. A pretrained personalized embedding is used to extract user-specific mutual influences between candidate items. 
PFRN \cite{pfrn} employs multiple self-attention structures to flight itinerary re-ranking. 
Instead of a simple concatenation of a pre-trained user representation with the candidate item representation as in PRM, PFRN exploits users' multiple behaviors, like long-term booking behaviors, real-time clicking behaviors, by individual multi-head self-attentions. The final prediction is generated by capturing the interactions between candidate items and users' multiple behaviors. 
A more recent work Raiss \cite{raise} attempts to improve personalization in re-ranking by maintaining individual attention weights in modeling cross-item interactions for each user.
% The authors also visualize the attention weights of their model to understand how candidate items influence each other. 

\smallskip
\noindent
\textbf{Other network structures.} 
To avoid potential position or contextual bias, List-CVAE \cite{listcvae} further explores conditional variational auto-encoders (CVAE) and directly learns the joint distribution of items conditioned on user responses. 
\citet{pivotcvae} find that the generative model of List-CVAE is usually trapped in a few items and fails to cover item variation in the re-ranking list. They propose a pivot selection phase (PivotCVAE) to improve the variation of the list.
HRM \cite{hrm} finds that introducing user behaviors and computing the similarity between candidate items and interested items in history improves the quality of re-ranking.
Liu \etal \shortcite{irgpr} further investigate the complementary and substitutable relationships among candidate items and propose a graph-based model, IRGPR.

Above we have provided a brief review of the development for methods of learning by observed signals. Most of the work formulates the re-ranking as a sequential modeling problem to extract the cross-item interactions on initial ranking lists. We also witness a trend of evolving from recurrent to self-attentive structures.
However, despite the various network structures, the above models are only trained with the only permutation that is displayed to the users, with other $n!-1$ permutations unexplored, limiting the potential of selecting the optimal permutation for re-ranking. In the next section, we will focus on learning by counterfactual signals that estimates listwise contexts on different permutations.

\subsection{Learning by Counterfactual  Signals}\label{sect:counterfactual}
 To provide signals on counterfactual lists, methods that learn by counterfactual signals usually follow \textit{an evaluator-generator paradigm} (EG)---with a generator to generate feasible permutations and an evaluator to evaluate the listwise utility of each permutation, as shown in Fig.~\ref{fig:flowchart}(b). 
 
%  The evaluator is trained to encode the listwise context of the re-ranked lists by the log data, and the generator is then supervised by the evaluator.

\citet{seg} first adopt the evaluator-generator paradigm and propose the SEG model. They point out two desired properties for an evaluator: (i) \textit{Order-sensitivity}, the evaluator needs to be sensitive to the order of the input lists; and (ii)\textit{ Generalizaility}, the evaluator shall generalize well to all possible permutations. 
Under the guidance of the evaluator, SEG devises a supervised learning approach and a reinforcement learning approach to train the generator. The supervised learning approach directly learns the estimated utility provided by the evaluator, while the reinforcement learning approach further pursues long-term reward in each step by the temporal difference (TD) error \cite{silver2014deterministic}.

A series of follow-up studies \cite{gcr,prs,grn,crum} explores various network structures like RNN or self-attention for evaluators and the generators. Specific structures are listed in Table \ref{tab:model_comparison}. Though the structures for the generator diverge, we observe that a common choice for the evaluator is the RNN-based structure, due to its satisfying performance in modeling users' sequential behaviors \cite{ncm}. 

In addition, these studies also focus on improving the training procedure of the generator. CDIA \cite{cdia} applies the actor-critic reinforcement learning and uses policy gradient with the advantage function to update the generator. To tackle the problem of large action space of $O(n!)$ for the generator, \citet{gcr} adapt the proximal policy optimization (PPO) algorithm and introduce PPO-exploration to train the generator in the proposed GCR model. PRS \cite{prs} exploits the beam search to generate feasible permutations and directly uses the evaluator to select the optimal list. GRN \cite{grn} also employs the policy gradient for optimization, whereas CRUM \cite{crum} utilizes Lambdaloss to train the generator for utility optimization. \citet{egrerank} notice that the evaluator is trained by the offline labeled data and may not generalize well to unseen distribution, and introduce EGRerank with a discriminator to provide a self-confidence score for the evaluation.

Two exceptions without the evaluator-generator architecture are GSF \cite{gsf} and SetRank \cite{setrank}, which learns permutation-invariant re-ranking models that are insensitive to permutations of the input.
A groupwise scoring function (GSF) \cite{gsf} is devised with DNN on all the  size-$m$ permutations of items in initial lists ($m\leq n$). 
\citet{setrank} apply a variant of self-attention structure without positional encoding and dropout (SetRank) to preserve the permutation invariant property.

Though potentially effective in selecting the optimal re-ranking list by modeling counterfactual permutations, the training procedure is often more complex and the performance depends greatly on the quality of the evaluator.

\begin{table}[t]
\centering
\caption{Comparison of accuracy-oriented re-ranking models. D/M/NP stands for personalized by input data/by model parameters/non-personalized, E/G represents evaluator/generator, $n$ is the re-ranking size, $h$ is the length of the user history, and $m$ is the permutation size for GSF.}
\label{tab:model_comparison}
 \begin{adjustbox}{max width=\linewidth}
% \begin{threeparttable}
\setlength{\tabcolsep}{1.6mm}
\begin{tabular}{lllll}
\toprule
                      & \begin{tabular}[c]{@{}l@{}}Listwise context \\ modeling\end{tabular}   & Optimization                     & P/NP               & Complexity                               \\\midrule
DLCM \shortcite{dlcm}                 & GRU                                                                    & AttRank                          & NP                 & $\mathcal O(n)$                                     \\
MiDNN\shortcite{midnn}                 & LSTM                                                                   & CE                               & D                  & $\mathcal O(n)$                                      \\
ListCVAE \shortcite{listcvae}             & CVAE                                                                   & KL                               & NP                 & $\mathcal O(n)$                                      \\
Seq2Slate \shortcite{seq2slate}            & PointerNet                                                             & CE                               & NP                 & $\mathcal O(n^2)$                   \\
HRM \shortcite{hrm}                  & Similarity                                                             & Hinge                            & D                  & $\mathcal O(hn+h^2)$                                     \\
PRM \shortcite{prm}                  & Self-attention                                                         & CE                               & D                  & $\mathcal O(n^2)$                                      \\
IRGPR  \shortcite{irgpr}               & GNN                                                                    & BPR                              & M                 & $\mathcal O(n)$                                      \\
PFRN \shortcite{pfrn}                 & Self-attention                                                         & CE                               & D                  & $\mathcal O(n^2+h^2)$                                     \\
PivotCVAE \shortcite{pivotcvae}             & CVAE                                                                   & KL                               & NP                 & $\mathcal O(n)$                                      \\
Raise \shortcite{raise}                & Self-attention                                                         & CE                               & M                  & $\mathcal O(n^2)$                                     \\
\midrule
 GSF \shortcite{gsf}                   & DNN                                                                    & CE                               & NP                 & $O\mathcal (\frac{mn!}{(n-m)!})$                                    \\
\multirow{2}{*}{SEG \shortcite{seg}}   & E: BiGRU                                                                 & \multirow{2}{*}{MSE/Q-learning}             & \multirow{2}{*}{D} & \multirow{2}{*}{$\mathcal O(n^2)$  }                    \\
                      & G: GRU                                                                 &                                  &                    &                                          \\
SetRank \shortcite{setrank}              & Self-attention                                                         & AttRank                          & NP                 & $\mathcal O(n^2)$                                       \\
\multirow{2}{*}{CDIA \shortcite{cdia}} & E: LSTM                                                                & \multirow{2}{*}{Policy gradient}       & \multirow{2}{*}{D} & \multirow{2}{*}{$\mathcal O(n^2)$} \\
                      & G: LSTM                                                                &                                  &                    &                                          \\
\multirow{2}{*}{GCR \shortcite{gcr}}  & E: BiGRU+attention                                                       & \multirow{2}{*}{PPO-exploration} & \multirow{2}{*}{D} & \multirow{2}{*}{$\mathcal O(n^2)$} \\
                      & G: GRU                                                                 &                                  &                    &                                          \\
\multirow{2}{*}{PRS \shortcite{prs}]}  & E: BiLSTM                                                                & \multirow{2}{*}{---}               & \multirow{2}{*}{D} & \multirow{2}{*}{$\mathcal O(n^2)$} \\
                      & G: Beam search                                                    &                                  &                    &                                          \\
\multirow{2}{*}{GRN \shortcite{grn}}  & E: BiLSTM+attention                                                      & \multirow{2}{*}{Policy gradient}       & \multirow{2}{*}{D} & \multirow{2}{*}{$\mathcal O(n^2)$} \\
                      & \begin{tabular}[c]{@{}l@{}}G: GRU+attention+\\ PointerNet\end{tabular} &                                  &                    &    
                      \\
\multirow{2}{*}{CRUM \shortcite{crum}}     & E: BiLSTM+GNN                                                          & \multirow{2}{*}{LambdaLoss}      & \multirow{2}{*}{D} & \multirow{2}{*}{$\mathcal O(n)$}                    \\
                          & G: MLP                                                               &                                  &                     &                                          \\
\multirow{2}{*}{EGRerank \shortcite{egrerank}} & E: LSTM                                                              & \multirow{2}{*}{PPO}             & \multirow{2}{*}{D}  & \multirow{2}{*}{$\mathcal O(n^2)$} \\
                          & G: LSTM                                                              &                                  &                     &                                                             \\
\bottomrule                                     
\end{tabular}
% \end{threeparttable}
\end{adjustbox}
\end{table}

\subsection{Qualitative Model Comparison}\label{sect:personalization}
Next, we give a thorough comparison of the above-mentioned models in terms of network structure, optimization, personalization, and computational complexity, as shown in Table~\ref{tab:model_comparison}. 

\smallskip
\noindent
\textbf{Network structure.} 
We notice that using self-attention, or a combination of RNN and the attention mechanism in re-ranking has been especially popular in recent years. For the design of the evaluator, bi-directional RNN (\eg, BiLSTM, BiGRU) is proved to be more effective in many studies \cite{seg,prs,crum}, where BiRNN is capable of capturing the two-way evolution of user’s interests during browsing \cite{grn}.

\smallskip
\noindent
\textbf{Optimization.}
For those models that learn by observed signals (the upper table), the loss function can be broadly grouped into pointwise (cross-entropy loss (CE)), pairwise (BPR loss \cite{rendle2012bpr}, hinge loss \cite{bartlett2008classification}), and listwise (Attention Rank loss (AttRank) \cite{dlcm}, KL loss). The pointwise CE loss is the most adopted loss due to its simplicity and effectiveness. 
Methods that learn by counterfactual signals often follow the evaluator-generator paradigm, where the training of the generator is guided by the evaluator. Since the ranking operation is discrete and non-differentiable, these models often rely on the policy gradient \cite{silver2014deterministic} or Lambdaloss \cite{wang2018lambdaloss} to optimize the model.

\smallskip
\noindent
\textbf{Personalization.}
Re-ranking results should be user-specific and cater to individual users' preferences and intents. Moreover, the cross-item interactions of item pairs vary from user to user. Thus personalization is an essential requirement for re-ranking. We observe a trend of emphasizing personalized models over non-personalized ones in recent years. There are mainly two ways to provide personalized re-ranking results: (i) \textit{personalization by input data} and (ii) \textit{personalization by model parameters}. The former way simply takes user features as input, \eg, the user profiles or the user historical behaviors, and extracts personal preferences by specific network architectures like self-attention in PFRN \cite{pfrn}. The network parameters are shared across users. While the latter maintains an individual set of parameters for each user as in Raiss \cite{raise} or IRGPR \cite{irgpr}. 
% \ruiming{why do we need to discuss personalization?}
% All users share the same parameters.

\smallskip
\noindent
\textbf{Complexity.}
Learning by observed signals directly predicts the re-ranking scores for $n$ items and are mainly of the linear time complexity $\mathcal O(n)$, except for the ones that apply the attentive listwise modeling. The runtime for the self-attention is quadratic in $n$ as it computes the interactions between any pair of items, but the calculation can be made parallel to accelerate the process \cite{vaswani2017attention}. Seq2Slate \cite{seq2slate} also yields a $\mathcal O(n^2)$ complexity---for each one of the $n$ steps, Seq2Slate examines the current remaining items and selects the best one from them.

% \footnote{The complexity for self-attention related models can be reduced to $\mathcal O(n)$ with  optimization in network structure \cite{shen2021efficient}.} (both for training and inference), with $n$ the re-ranking size. The only exception is Seq2Slate \cite{seq2slate}. , thus yields a $\mathcal O(n^2)$ complexity. HRM \cite{hrm} and PFRN \cite{pfrn} compute the interactions between the candidate items and items in the users' history list. Suppose the length of the users' history is $h$, the complexity is, therefore, $\mathcal O(hn)$.

As for learning by counterfactual signals, GSF takes $m$-permutations of $n$ items as input so that the complexity is $O\mathcal (\frac{mn!}{(n-m)!})$. For most evaluator-generator models, the generators sequentially select the next item similar to Seq2Slate, leading to a polynomial complexity $\mathcal O(n^2)$. CRUM \cite{crum}, on the other hand, though have a $\mathcal O(n^2)$ training time, the time complexity for inference is $\mathcal O(n)$ by directly predicting the re-ranking scores for all the $n$ items with an MLP structure. 

% the complexity is generally higher. For example,

\section{Multiple Objectives}\label{sect:multiple}
Accuracy is no doubt the most important objective for recommender systems. Apart from accuracy, other objectives like diversity or fairness are also crucial measurements in MRS. Purely optimizing accuracy, if applied carelessly, can yield similar or near-duplicate results and further result in the \textit{echo chamber} effects \cite{ge2020understanding}. Many studies aim to simultaneously optimize accuracy and other objectives (diversity/fairness). How to delicately manage the tradeoff between multiple objectives becomes a key problem, as sometimes diversity and fairness can contradict accuracy \cite{liu2019personalized}. We introduce diversity-aware and fairness-aware re-ranking in Section \ref{sect:div} and \ref{sect:fair}, respectively.

\subsection{Diversity-aware Re-ranking}\label{sect:div}

Diversity usually measures the dissimilarity of the re-ranking list for each user. In contrast to non-learning re-ranking methods like MMR \cite{carbonell1998use}, neural diversity-aware models usually conduct an end-to-end learning scheme, with no need of handcrafting relevance and diversity features. Below we give a brief review of neural diversity-aware re-ranking by broadly classifying existing studies into \textit{implicit approaches} and \textit{explicit approaches}. The implicit approaches measure diversity by inter-item similarity and do not require subtopics (\eg, category of items) to evaluate diversity, while explicit approaches aim at promoting the coverage of items over specified subtopics. 
% \ruiming{implicit/explicit or item/subtopic level?}

For implicit approaches, NTN \cite{ntn} proposes a neural tensor network to learn the dissimilarity between any pairs of items. The re-ranking list is sequentially generated by a linear combination of the relevance and the dissimilarity of candidate items. MDP-DIV \cite{mdpdiv} directly optimizes general diversity measures like $\alpha$-DCG or $S$-recall, and uses the policy gradient to optimize the long-term reward. M$^2$DIV \cite{m2div} enhances MDP-DIV by introducing LSTM and the lookahead Monte Carlo Tree Search (MCTS) to the ranking policy.  \citet{daletor} derive a smooth approximation of diversity metrics in the proposed DALETOR model and apply a self-attention structure to model the listwise context.

While for explicit approaches, \citet{dssa} notice the advantage of using the attention mechanism to determine the importance for the under-covered subtopics, and propose DSSA, where the relevance and the diversity are jointly estimated with a subtopic attention.
As a follow-up to DSSA, DVGAN \cite{dvgan} formulates the problem of generating diverse re-ranking lists as a minimax game. It adapts DSSA as a generator, and involves a discriminator to determine how relevant and diverse the given list is. DESA \cite{desa} explores to leverage item dependencies in terms of both relevance and diversity, which is composed of an encoder and a decoder with the self-attention to extract item and subtopic correlations. \citet{airbnb} investigate the potential of deploying a diversity-aware re-ranking to Airbnb search. They design a metric for measuring the distance between two lists and use an LSTM structure to generate the re-ranking list.

The balance between accuracy and fairness is managed either by learning a trade-off parameter \cite{ntn}, or directly optimizing a specific metric that combines accuracy and fairness like $\alpha$-NDCG \cite{daletor}. 

% The complexity of diversity-aware re-ranking, except for DESA of linear complexity, are all $\mathcal O(n^2)$ as the diversity is measured between the remaining candidates and the previously selected items at each step to sequentially generate the next item. 
% Approaches above only model the cross-item interactions with regard to diversity, while

\subsection{Fairness-aware Re-ranking}\label{sect:fair}
Fairness, with a growing influence on IR community, has been made a critical objective for re-ranking. In this review, we focus on the \textit{item fairness}, since it is the main focus of existing re-ranking literature. Item fairness ensures each item or item group receives a fair proportion of exposure (\eg, proportional to its merits or utility). Neural re-ranking, however, has been a relatively under-explored domain. FMMR \cite{fmmr} first constructs fairness representation for each demographic group using CNN and adopts MMR to trade-off relevance and fairness. \citet{pgrank} aim to optimize a general utility metric while satisfying the fairness of exposure constraints by the Plackett-Luce model \cite{plackett1975analysis} in PGRank. A follow-up study, PLRank \cite{plrank}, improves the policy gradient in PGRank by deriving an unbiased estimate of the gradient. FULTR \cite{fultr} further explores a counterfactual estimate for both utility and fairness constraints for the Plackett-Luce model. \citet{gen} empirically show the prevalence of unfairness in cold-start recommendation, and propose an auto-encoder re-ranking model, GEN, to alleviate the fairness issue for cold-start items. 

\begin{table*}[]
\centering
% \footnotesize
\caption{Performance Comparison on Ad and PRM Public datasets. The initial ranking list (Init) is produced by LambdaMart.}
\label{tab:overall}
\setlength{\tabcolsep}{5pt}
\renewcommand{\arraystretch}{0.8}
\begin{adjustbox}{max width=0.95\linewidth}
\begin{tabular}{l cccc cccc}
\toprule
\multirow{2}{*}{} & \multicolumn{4}{c}{Ad} & \multicolumn{4}{c}{PRM Public} \\
\cmidrule(lr){2-5} \cmidrule(lr){6-9}
 & MAP@5 & NDCG@5 & MAP@10 & NDCG@10 & MAP@10 & NDCG@10 & MAP@20 & NDCG@20 \\
 \midrule
Init \shortcite{lambdamart} & 0.6037 & 0.6840 & 0.6075 & 0.6990 & 0.1842 & 0.2178 & 0.1901 & 0.3202 \\ \midrule
MiDNN \shortcite{midnn} & 0.6080 & 0.6876 & 0.6117 & 0.7021 & 0.3069 & 0.3482 & 0.2977 & 0.4265 \\
GSF \shortcite{gsf} & 0.6090 & 0.6883 & 0.6126 & 0.7028 & 0.3060 & 0.3459 & 0.2968 & 0.4241 \\
EGRerank \shortcite{egrerank} &	0.6092 & 0.6890 & 0.6126 & 0.7029 &	0.3075 & 0.3502 & 0.2985 & 0.4286\\
DLCM \shortcite{dlcm} & 0.6126 & 0.6914 & 0.6162 & 0.7055 & 0.3082 & 0.3500 & 0.2991 & 0.4287 \\
SetRank \shortcite{setrank} & 0.6132 & 0.6917 & 0.6168 & 0.7060 & 0.3094 & 0.3515 & 0.3002 & 0.4297 \\
PRM \shortcite{prm} & 0.6140 & 0.6923 & 0.6178 & 0.7066 & 0.3096 & 0.3516 & 0.3003 & 0.4301\\
 \bottomrule
\end{tabular}
\end{adjustbox}
\end{table*}

\section{Emerging Applications}\label{sect:app}
Neural re-ranking has also been seen in many emerging and interesting industrial applications.

\subsubsection{Integrated Re-ranking}\label{sect:int}
Integrated re-ranking (\textit{a.k.a.,} mixed re-ranking) is a rapidly emerging domain driven by practical problems, where the MRS is required to display \textit{a mix of items} from different sources/channels with heterogeneous features \eg, integrated feeds of articles, videos, and news \cite{hrlrec}. The input is extended from a single list to multiple lists.
DHANR \cite{dhanr} proposes a hierarchical self-attention structure to consider cross-channel interactions.
\citet{hrlrec} decompose the integrated re-ranking problem into two sub-tasks---source selection and item ranking, and use hierarchical reinforcement learning (HRL) to solve the problem. DEAR \cite{dear,ram} learns to interpolate ads and organic items by the designed deep Q-networks. \citet{liao2021cross} also adopts a reinforcement learning solution with a cross-channel attention unit.

\subsubsection{Edge Re-ranking}\label{sect:edge}
In a framework of \textit{cloud-to-edge}, \citet{gong2020edgerec} find that real-time computing on edge helps capture user preferences more delicately and improve the performance of recommendations. Therefore, they propose EdgeRec, which generates initial ranking lists on cloud, and conducts re-ranking with instant feedback on mobile devices. Edge re-ranking opens up interesting research topics especially for on-device personalized models or federated learning \cite{hard2018federated}.

\section{Experiments}
For understanding and analyzing the performance of re-ranking algorithms, we provide a re-ranking library---\texttt{LibRerank}, which automates the re-ranking experimentation and integrates a major collection of re-ranking algorithms. It is designed to support researchers by simplified access to popular re-ranking algorithms, thereby making experimental results more reproducible.

We conduct benchmarking experiments on two public recommendation datasets, \textbf{Ad}\footnote{https://tianchi.aliyun.com/dataset/dataDetail?dataId=56} and \textbf{PRM Public}\footnote{https://github.com/rank2rec/rerank}. A detailed explanation of the benchmarking experiments, and the processed datasets are also released together with the \texttt{LibRerank} library\footnote{https://github.com/LibRerank-Community/LibRerank}. We use LambdaMART \cite{lambdamart} to produce the initial ranking lists. Baselines include: MiDNN \cite{midnn}, GSF \cite{gsf}, DLCM \cite{dlcm}, PRM \cite{prm}, SetRank \cite{setrank}, and EGRerank \cite{egrerank}. We anticipate adding support for more re-ranking algorithms, including diversity- or fairness-aware ones in the near future. 
% \ruiming{any reason?}

\smallskip
\noindent
\textbf{Principles.}
For fair comparisons, our implementation follows several principles: (i) To cover every detail in each algorithm, we use the open-sourced implementation if applicable. Otherwise, the algorithms are implemented according to the original paper. (ii) We conduct careful parameter tuning for every algorithm and report the best results. 

\subsubsection{Quantitative Evaluation}
For the quantitative evaluation, we focus on the popular ranking metrics MAP@$k$ and NDCG@$k$, with $k=5,10$ for Ad, $k=10,20$ for PRM Public due to the different re-ranking sizes. The results are reported in Table \ref{tab:overall}, from which we have the following observations.

\noindent
\textit{(i) Effectiveness of Re-ranking.} The first row in Table \ref{tab:overall} shows the performance of the initial ranking, generated in the ranking stage by LambdaMART. The results of all the re-ranking algorithms are appealing and outperform the initial ranker by a large margin. This confirms the necessity of the re-ranking stage in MRS by integrating the listwise context.

\noindent
\textit{(ii) Listwise Context Modeling.} Considering re-ranking algorithms with different listwise context modeling structure, algorithms with self-attention architecture like SetRank and PRM, achieves better results. It is because the self-attention structure effectively encodes the cross-item interactions between any pairs of items.

\noindent
\textit{(iii) Robustness of EG framework.} EGRerank adopts an evaluator-generator (EG) paradigm, but its performance is less impressive. Possible reasons may be that the performance of the generator largely depends on the quality of the evaluator, which is relatively hard to measure and select.

% while it is hard to measure and select a good evaluator.

% We also note that the differences between re-ranking algorithms are relatively small in Table \ref{tab:overall}. It is because the datasets are extremely sparse, with an average of only one positive item over the whole list. Putting the positive item to different positions yields quite similar MAPs and NDCGs. While the reported improvements are all significant compared with the second-best baselines.

% \subsubsection{Re-ranking Size}
% The performance of re-ranking models may depend on the size of the item list to be re-ranked. The re-ranking size is usually 10-50, depending on the latency requirement of the system. For practical considerations on computing efficiency, the re-ranking models usually do not rank too many items.

% \subsection{Ranking and Re-ranking}
% The performance of the ranking models also influence the re-ranking stage.

% \subsection{Code}
% \subsection{How the re-ranking models depend on the performance of the ranking models.}
\section{Summary and Future Prospects}
Over the past several years, neural re-ranking has continued to become an inspiring domain, motivated by both scientific challenges and industrial demands. A considerable amount of studies have been conducted, and many of them have already found use in industrial applications. Major advances in this domain are summarized in Fig.~\ref{fig:orgchart}. A review of the re-ranking algorithms with corresponding objectives can be found in Section \ref{sect:acc} and \ref{sect:multiple}. Our benchmarking results manifest the superiority of the neural re-ranking models. Despite the great progress in recent years, we still note that there are some significant challenges and open issues in this domain. 

\smallskip
\noindent
\textbf{Sparse Feedback.}
The re-ranking problem is challenging due to the sparse supervised signal, where only the feedback for the displayed lists can be observed---feedback for the other $n!-1$ permutations is unavailable. Evaluators or click models \cite{ncm,cacm,ccm} can be potentially used to generate feedback, but current click models are just trained to fit the offline click data by performance metrics like log-likelihood, without particular designs on how evaluators should be trained to address the data sparsity problem and help improve the training of the re-ranking models.

\smallskip
\noindent
\textbf{Personalization for Diversity/Fairness}.
Personalization is the core of MRS, but recent literature mostly focuses on personalization in accuracy-oriented re-ranking, leaving personalization in diversity and fairness unexplored. Different users have various demands for diversity and fairness. 
It is of great potential to involve personalized diversity or fairness. % concerns.

\smallskip
\noindent
\textbf{Tradeoff between Multiple Objectives.}
Different recommendation scenarios have different degrees of demand for diversity or fairness. Existing studies mainly manage the tradeoff by heuristics or parameter tuning. It could be a promising topic to automatically balance multiple objectives without human intervention.

\smallskip
\noindent
\textbf{Joint Training of MRS.}
Re-ranking models are trained separately, decoupling from other stages in MRS. But the ranking quality of other stages affects the performance of re-ranking \cite{seq2slate}. Utilizing the information learned by other stages (\eg, parameter transfer, gradient transfer) would be of high value for both academia and industry.

\newpage
\bibliographystyle{named}
\bibliography{shortref}

\end{document}